 \documentclass[%
 reprint,%
 amssymb, amsmath,%
 aip,cha,%
 groupedaddress,%
 frontmatterverbose,
 ]{revtex4-1}
 \usepackage{graphicx}
 \usepackage{bm}
 \usepackage{bbm}

 \usepackage{subfigure}
 \usepackage{amssymb}
 \usepackage{xcolor}

 \usepackage{xcolor,xspace,colortbl,rotating}
 \usepackage{amsfonts}
 \usepackage{tabulary}

\begin{document}
\title{Manipulating antiferromagnets with magnetic fields: ratchet motion of multiple domain walls induced by asymmetric field pulses}

\author{O. Gomonay}
\affiliation{Institut f\"ur Physik, Johannes Gutenberg Universit\"at Mainz, D-55099 Mainz, Germany}
\affiliation{National Technical University of Ukraine ``KPI'', 03056, Kyiv,
	Ukraine}
\author{M. Kl\"aui}
\affiliation{Institut f\"ur Physik, Johannes Gutenberg Universit\"at Mainz, D-55099 Mainz, Germany}
\author{J. Sinova}
\affiliation{Institut f\"ur Physik, Johannes Gutenberg Universit\"at Mainz, D-55099 Mainz, Germany}
\affiliation{Institute of Physics, Academy of Sciences of the Czech Republic, Cukrovarnicka 10, 162 00 Praha 6, Czech Republic}

\begin{abstract}

	Future applications of antiferromagnets (AFs) in many spintronics devices rely on the precise manipulation of domain walls. The conventional approach using static magnetic fields is inefficient due to the low susceptibility of AFs. Recently proposed electrical manipulation with spin-orbit torques is restricted to metals with a specific crystal structure. 
 Here we propose an alternative, broadly applicable approach: using  asymmetric magnetic field pulses  to induce controlled  ratchet motion of AF domain walls. 
 The efficiency of this approach is based on three peculiarities  of AF dynamics. First,  a time-dependent magnetic field couples with an  AF order parameter  stronger  than a static magnetic field, which leads to higher mobility of the domain walls. Second,  the rate of change of the magnetic field couples with the spatial variation of the AF order parameter inside the domain and this  enables synchronous motion of multiple domain walls with the same structure. Third, tailored asymmetric field pulses in combination with static friction can prevent backward motion of domain walls and thus lead to the desired controlled ratchet effect. 
 The proposed use of an external field, rather than internal spin-orbit torques, avoids any restrictions on size, conductivity, and crystal structure of the AF material. 
 We believe that our approach paves a way for the development of new AF-based devices based on controlled motion of AF domain walls. 
\end{abstract}

\maketitle



Antiferromagnets (AFs) are considered  perspective materials for spintronic applications: they exhibit fast magnetic dynamics with excitations in the THz range, are fundamentally insensitive to external magnetic fields,  and produce no stray fields.\cite{MacDonald2011, Gomonay2014, Jungwirth2016} One of the further advantages of AFs, important for fast switching between different states, is related with the motion of domain walls (DWs). In contrast to ferromagnets (FM), the dynamics of AF DWs shows no Walker breakdown.
 Thus, the DW velocity is only limited by the group velocity of spin waves, which is of the order of tens of km/s (e.g., 40 km/s for NiO). This is orders of magnitude larger than the typical velocities in FM, where the Walker breakdown limits the achievable velocities, and also larger than velocities in synthetic AFs. \cite{Yang2015a}

However, the manipulation of AF DWs faces significant difficulties. In particular, 180$^\circ$ AF domains are indistiguishable even in the  presence of a constant homogeneous magnetic  field. So, in contrast to FMs, an applied external field cannot move the 180$^\circ$ AF DWs at all. In addition,  coupling between the external magnetic field and the AF order parameter (N\'eel vector) is  suppressed due to the strong exchange coupling between the magnetic sublattices. In this case typical values of fields necessary to produce any noticeable shift of the DW  are of the order of the spin-flop field and range 1-10 T. \cite{Barthem2016}

Recently the possibility to move DW in an AF with the help of a staggered N\'eel spin-orbit torque was demonstrated in Ref. \onlinecite{Gomonay2016}. While this mechanism can be very effective, its application is restricted to metals  that have a broken local inversion symmetry, which the vast majority of the AF systems do not have. Furthermore,  manipulation using regular spin-orbit torques has been shown to be 
restricted to specific DW types, sample geometry and AF spin structure configuration, which narrows the applicability of these torques.\cite{Shiino2016}

Finally, recent calculations predict that temperature gradients  can move the AF DWs in metals and isolators as well.\cite{Kim2015d,Selzer2016}  However, manipulation of the DWs using this mechanism is restricted to one-directional motion and is yet to be observed.
Hence, at present there is no  broadly applicable  approach to manipulate  AF DWs.

In this Letter we develop such a broadly applicable approach to manipulate  AF DWs based on the use of asymmetric magnetic field pulses. We show that this approach is highly efficient for devices as it enables to attain high DW mobilities, to induce synchronous motion of multiple DWs and to control the DW displacement through a ratchet effect. 

We compare the dynamics of AF DWs induced by static and by time-dependent magnetic fields and show that a time-dependent field produces a larger effective force than its static counterpart. This difference originates from the strong exchange field  which reduces the magnetic static susceptibility. 
Our results show that the force produced by the rate of change of the magnetic field will move DWs with similar structure in the same direction.  
In contrast, the force produced by a static magnetic field is independent of the DW structure and  induces 
a shrinking and disappearance of unfavourable domains.  
We find the conditions for ratchet-like motion by calculating the critical rate of magnetic field that overcomes a static friction  force. 
We also propose an optimal configuration to implement controlled DW motion for the archetypical AFs Mn$_2$Au and NiO.

\begin{figure}[h]
	\centering
	\includegraphics[width=0.8\columnwidth]{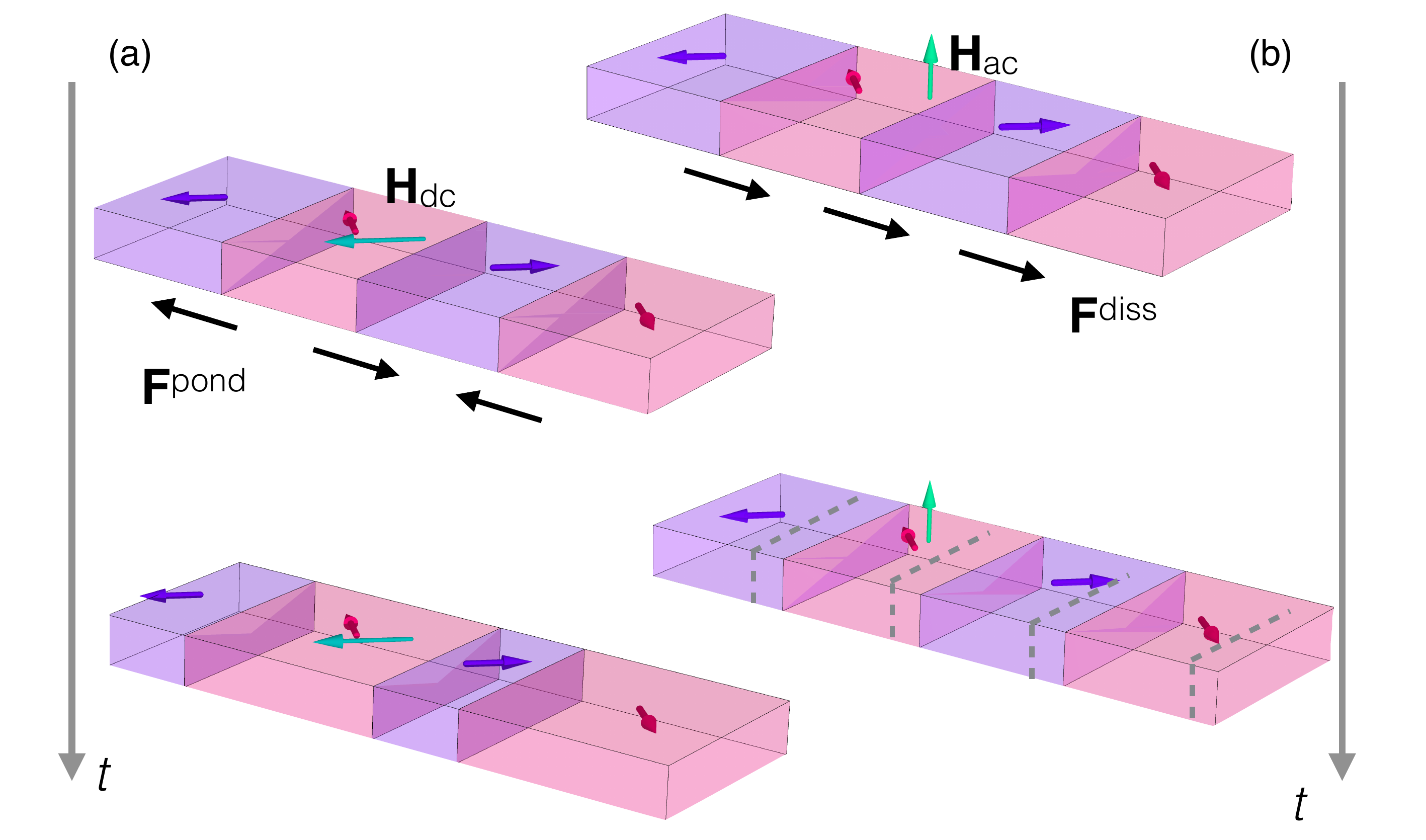}
	\caption[fig_stripe]{Evolution of the stripe AF domain structure induced by a constant, $\mathbf{H}_\mathrm{dc}$, (a) and a time-dependent, $\mathbf{H}_\mathrm{ac}$, (b) magnetic field. Black arrows show the directions of the ponderomotive, $\mathbf{F}^\mathrm{pond}$, and the dissipative, $\mathbf{F}^\mathrm{diss}$, forces. The grey dashed lines in (b) mark the previous position of the DWs.}
		\vspace{-0.2 cm}
	\label{fig_stripe}
\end{figure}

We consider the generic case of a compensated AF with two magnetic sublattices with magnetizations $\mathbf{M}_1$ and $\mathbf{M}_2$ ($|\mathbf{M}_1|=|\mathbf{M}_2|=M_s/2$). The metallic Mn$_2$Au and the isolating NiO are good examples of such AFs. 

The magnetic structure of an AF texture can be explicitly decribed in terms of the AF (N\'eel) vector $\mathbf{L}=\mathbf{M}_1-\mathbf{M}_2$ which is considered as a field variable, $\mathbf{L}(\mathbf{r},t)$.  The  closed equations of motion for AF vector \cite{Haldane1983a, Kosevich1990, Ivanov1995} have the following form:\cite{Gomonay2010}
\begin{equation}\label{eq_motion_AF_ideal}
\mathbf{L}\times\left[\ddot{\mathbf{L}}-c^2\Delta\mathbf{L} +\gamma^2H_\mathrm{ex}M_s\frac{\partial w_\mathrm{AF}}{\partial\mathbf{L}}\right]=\mathbf{T}-\gamma\alpha_GH_\mathrm{ex}\mathbf{L}\times\dot{\mathbf{L}}.
\end{equation}
In Eqs.~(\ref{eq_motion_AF_ideal})  we introduced the magnon velocity $c$  which coincides with the limiting velocity for the DW motion, $\gamma$ is the gyromagnetic ratio, and $w_\mathrm{an}$ is the density of magnetic anisotropy energy which depends upon the crystal structure. The effective field  $H_\mathrm{ex}$ parametrizes the exchange coupling between the magnetic sublattices. The last term in the r.h.s. of Eq.~(\ref{eq_motion_AF_ideal}) describes viscous damping parametrized by the Gilbert  constant $\alpha_G$. 

The vector $\mathbf{T}$ in the r.h.s. of Eqs.~(\ref{eq_motion_AF_ideal}) describes the effective forces (torques) induced by the external magnetic field $\mathbf{H}$:
\begin{eqnarray}
\mathbf{T}=
\gamma\mathbf{L}\times\dot{\mathbf{H}}\times\mathbf{L}-2\gamma \dot{\mathbf{L}}(\mathbf{H}\mathbf{L})-\gamma^2\mathbf{L}\times \mathbf{H} \left (\mathbf{L}
\cdot \mathbf{H} \right ).
\end{eqnarray}

In many practical cases the shape of the moving DW does not change or changes  slightly. So, the DW can be considered as a point particle, whose dynamics is described by only two vectors: the generalized momentum $\mathbf{P}$ and its canonically conjugated coordinate $\mathbf{R}$  (position of the DW center). The dynamics of Eq.~(\ref{eq_motion_AF_ideal}) can then be reduced to a standard equation for a point mass: \cite{Kosevich1990}
\begin{equation}\label{eq_point-mass_equation}
\frac{d\mathbf{P}}{dt}=-\gamma\alpha_GH_\mathrm{ex}\mathbf{P}+\mathbf{F},
\end{equation}
where $\mathbf{F}$ is the resulting external force, and the first term in the r.h.s. is analogous to viscous damping with relaxation time  $\tau_\mathrm{relax}=1/(\gamma\alpha_GH_\mathrm{ex})$.

Equation (\ref{eq_point-mass_equation}) is derived from the original Eq.~(\ref{eq_motion_AF_ideal}) in the following way.  First, we define the DW momentum $\mathbf{P}$ as an integral of  motion related with homogeneity of space:
\begin{equation}\label{eq_canonical_momentum}
P_j=-\frac{1}{\gamma^2M_sH_\mathrm{ex}}\int \dot{\mathbf{L}}^{(0)}\partial_j{\mathbf{L}^{(0)}}dV,\,j=x,y,z.
\end{equation}
Here $\mathbf{L}^{(0)}(\mathbf{r},t)$ is a solution of Eq.~(\ref{eq_motion_AF_ideal}) in the absence of a field ($\mathbf{T}=0$) and damping ($\alpha_G=0$). Second, we assume that $\mathbf{L}(t,\mathbf{r})=\mathbf{L}^{(0)}(t,\mathbf{r}-\mathbf{R})$. Finally, calculating explicitly the time derivative of Eq.~(\ref{eq_canonical_momentum}) and  taking into account Eq.~(\ref{eq_motion_AF_ideal}) we obtain Eq.~(\ref{eq_point-mass_equation}).


Among the forces, acting on the DW, we specify three types, essential for our consideration, $\mathbf{F}=\mathbf{F}^\mathrm{pond}+\mathbf{F}^\mathrm{diss}+\mathbf{F}^\mathrm{fric}$. The first one is the ponderomotive force
\begin{equation}\label{eq_ponderomotive force}
\mathbf{F}^\mathrm{pond}=\frac{\mathbf{n}S}{2 M_{s} H_\mathrm{ex}} \left[\left (\mathbf{L}_2
\mathbf{H}\right )^{2}-\left (\mathbf{L}_1
\mathbf{H}\right )^{2}\right].
\end{equation}
It stems from the difference in energy density between the left (N\'eel vector $\mathbf{L}_1$) and right (N\'eel vector $\mathbf{L}_2$) AF domains and is directed along the 
normal to the DW plane, $\mathbf{n}$ (see, e.g. Fig.~\ref{fig_stripe}). The ponderomotive force is proportional to the square of the magnetic field. Its value is weakened due to the strong exchange coupling between the magnetic sublattices. In addition, this force is insensitive to the structure of the DW itself and acts equally on the Bloch-like and N\'eel-like DW.

\begin{figure}[h]
	\centering
	\includegraphics[width=1.0\columnwidth]{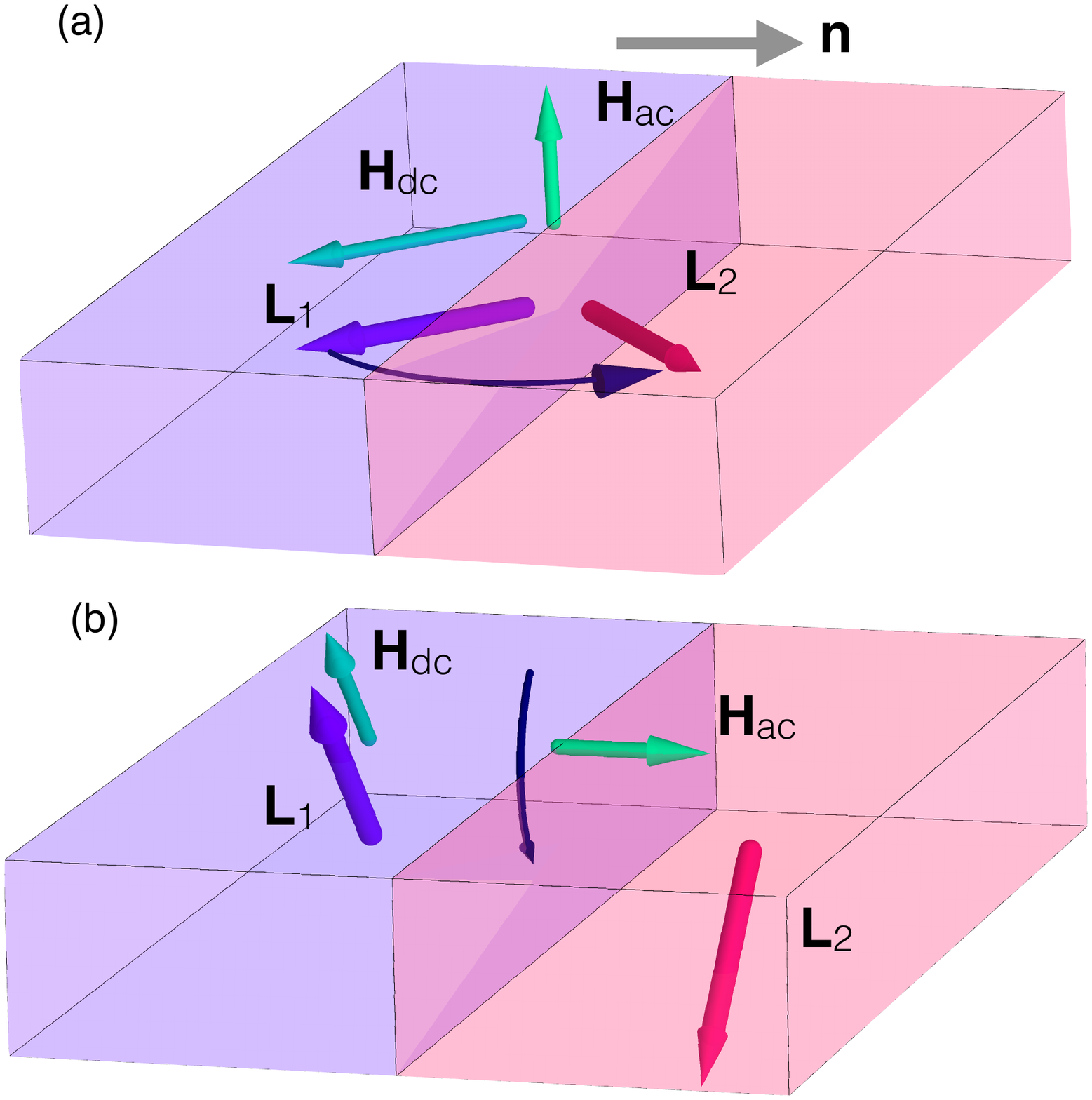}
	\caption{Sample with a single N\'eel (a) or Bloch (b) domain wall and optimal orientation of static ($\mathbf{H}_\mathrm{dc}$) and time-dependent ($\mathbf{H}_\mathrm{ac}$)  field. Curved arrows show the direction in which the AF vector rotates within the DW.}
	\label{fig_sample}
\end{figure}

The second force is dissipative and it is given by
\begin{equation}\label{eq_dissipative}
\mathbf{F}^\mathrm{diss}\cdot\mathbf{n}=-	\frac{1}{\gamma M_sH_\mathrm{ex}}\int\dot{\mathbf{H}}\cdot\mathbf{L}^{(0)}\times(\mathbf{n}\cdot\nabla)\mathbf{L}^{(0)}dV.
\end{equation}
It is induced by the time-dependent component of the magnetic field and is sensitive to the relative orientation of the the external field and AF vectors inside the DW, i.e. the DW structure. This force is maximal if the magnetic field is perpendicular to the $(\mathbf{L}_1,\mathbf{L}_2)$ plane, see, e.g. Fig.~\ref{fig_sample}. 
In spite of the small factor $1/H_\mathrm{ex}$, the dissipative force can be larger than $\mathbf{F}^\mathrm{pond}$, especially for high frequencies. Moreover, in contrast to $\mathbf{F}^\mathrm{pond}$, the dissipative force can move 180$^\circ$ domain walls.

\begin{figure}[h]
	\centering
	\includegraphics[width=0.7\columnwidth]{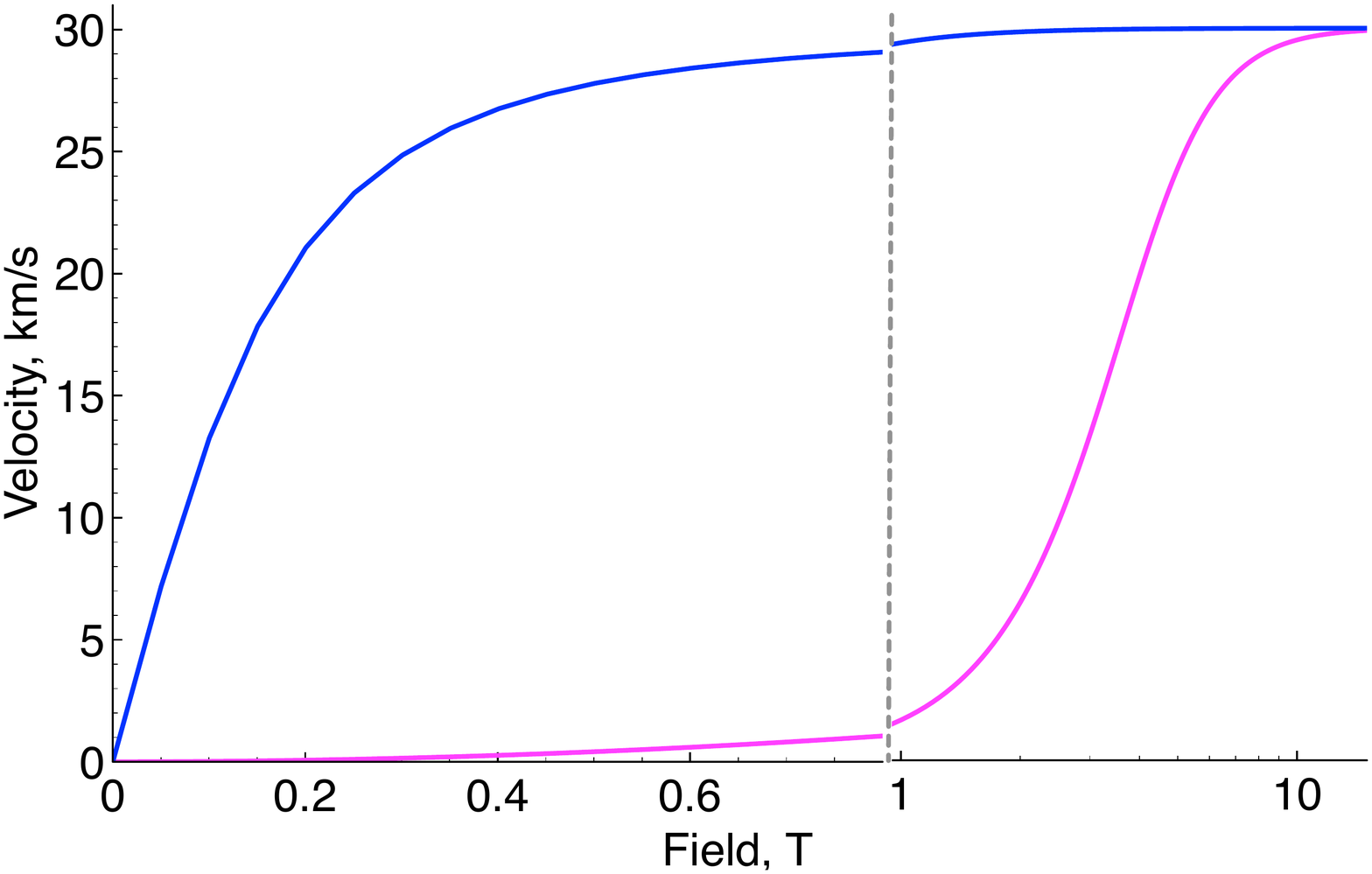}
	\caption[fig3]{Velocity of a $90^\circ$ AF DW vs static field  (magenta line) and an amplitude of steadily increasing time-dependent field  (blue line)  $H_\mathrm{ac}(t)=H^{(0)}_\mathrm{ac}t/\tau_\mathrm{raise}$  calculated according to Eq.~(\ref{eq_velocity of steady motion_parallel}) for Mn$_2$Au,  $\tau_\mathrm{raise}=5$ ps.  The vertical dashed line separates regions with linear and log-scale. }
	\label{fig_steady_velocity}
\end{figure}

Lastly, the third force is a friction force, $\mathbf{F}^\mathrm{fric}$. It is related to the magnetic defect distribution within the crystal and pinning strength of the defects. We consider this force as a static friction force which defines a threshold for the dynamics. This force is sample-dependent and can be estimated from the coercitivity. In the calculations below we 
take it to be 10\% of the spin-flop field.  

The important difference between the ponderomotive and dissipative forces is illustrated in Fig.~\ref{fig_stripe}, where we consider a stripe AF domain structure. The ponderomotive force  is directed from the favourable domain (with lower energy density) to the unfavourable. So, in a stripe structure adjacent DWs move in opposite directions, thus shrinking the fraction of unfavourable domains. Contrary to this,  the orientation of the dissipative force depends upon the AF DW structure. So, all the DWs with the same chirality  move in the same direction. Thus, the application of a time-dependent field provides an effective tool for manipulating AF-domains in an AF-based race-track type memory. 

We next analyse the dynamics of AF DWs through Eq.~(\ref{eq_point-mass_equation}).
We consider the simple case of a tetragonal AF (e.g. Mn$_2$Au) and 90$^\circ$ domain structure with orthogonal AF vectors in neighboring domains,  $\mathbf{L}_1\perp\mathbf{L}_2$. In this case the optimal orientation of the static magnetic field, $\mathbf{H}_\mathrm{dc}$, which produces the ponderomotive force, is parallel to one of the N\'eel vectors, e.g. $\mathbf{H}_\mathrm{dc}\|\mathbf{L}_1$ (see Fig.~\ref{fig_sample}). On the other hand, the optimal orientation of the time dependent field, $\mathbf{H}_\mathrm{ac}$, is related to the DW type.


In a thin film the AF vectors inside the DW rotate within the film plane and the DW is of a N\'eel type. For this case, the most efficient $\mathbf{H}_\mathrm{ac}$ is perpendicular to the film plane (Fig. ~\ref{fig_sample}a). In a bulk sample, a Bloch wall is also possible and  the most efficient $\mathbf{H}_\mathrm{ac}$ is perpendicular to the DW plane  (Fig. ~\ref{fig_sample}b). In both geometries $\mathbf{H}_\mathrm{ac}\perp\mathbf{H}_\mathrm{dc}$ and thus the time-dependent component does not contribute to the ponderomotive force.

Although $\mathbf{H}_\mathrm{dc}$ and $\mathbf{H}_\mathrm{ac}$ fields have different orientations, the corresponding forces, $\mathbf{F}^\mathrm{pond}$ and $\mathbf{F}^\mathrm{diss}$, are both parallel to the DW normal.

The time dependent component of the magnetic field allows one to manipulate the AF DW motion in a very effective way. To illustrate this fact, we start from  the constant (time-independent) forces produced by $\mathbf{H}_\mathrm{dc}$ and  steadily increasing/decreasing $\mathbf{H}_\mathrm{ac}$.  The velocity of the steady motion is
\begin{equation}
v_\mathrm{steady}=c\frac{(\pi\dot{H}_\mathrm{ac}/\gamma +H_\mathrm{dc}^{2})/(2H_{\mathrm{ex}})}{\sqrt{\alpha _{G}^{2}H_{\mathrm{an}}H_{%
			\mathrm{ex}}+(\pi\dot{H}_\mathrm{ac}/\gamma-H_\mathrm{dc}^{2})^2/(2H_{\mathrm{ex}})^{2}}},
\label{eq_velocity of steady motion_parallel}
\end{equation}%
as can be obtained from Eq.~(\ref{eq_point-mass_equation}). Here $H_{\mathrm{an}}$ is the anisotropy field.


Contributions of the time-dependent and the static component to $v_\mathrm{steady}$ are compared in Fig.~\ref{fig_steady_velocity}. For the calculations we use field values  $H_\mathrm{ex}$=1400 T,  $H_\mathrm{an}$=30 mT typical for  AFs with high N\'eel temperature (like Mn$_{2}$Au \cite{Wu2012, Shick2010} and NiO \cite{Hutchings1972}). We set the AF magnon velocity $c=30$ km/s. As the damping parameters of metals and insulators are different,  we take $\alpha_G=10^{-4}$ for insulating NiO \cite{Kampfrath2010} and $\alpha_G=10^{-3}$ for metalic Mn$_{2}$Au. These values correspond to relaxation times $\tau_\mathrm{relax}=$ 50 ps and 5 ps, respectively. The friction force per unit DW area is taken 9 N/m$^2$ which corresponds to an effective coercive field of 0.1 T. 


 Fig.~\ref{fig_steady_velocity} shows that the mobility ($=dv/dH$) of the DW in an ac field is much higher than in the static field and an amplitude value ${H}_\mathrm{ac}=$1 T is enough to reach the limiting velocity. 
 However, a practical fast increase of the magnetic field is only possible on short time scales and up to a limited amplitude of  $H_\mathrm{ac}$. These facts exclude monotonously varying $\mathbf{H}_\mathrm{ac}$ as a useful tool for DW manipulation.
 
\begin{figure}[h]
	\centering
	\includegraphics[width=0.9\columnwidth]{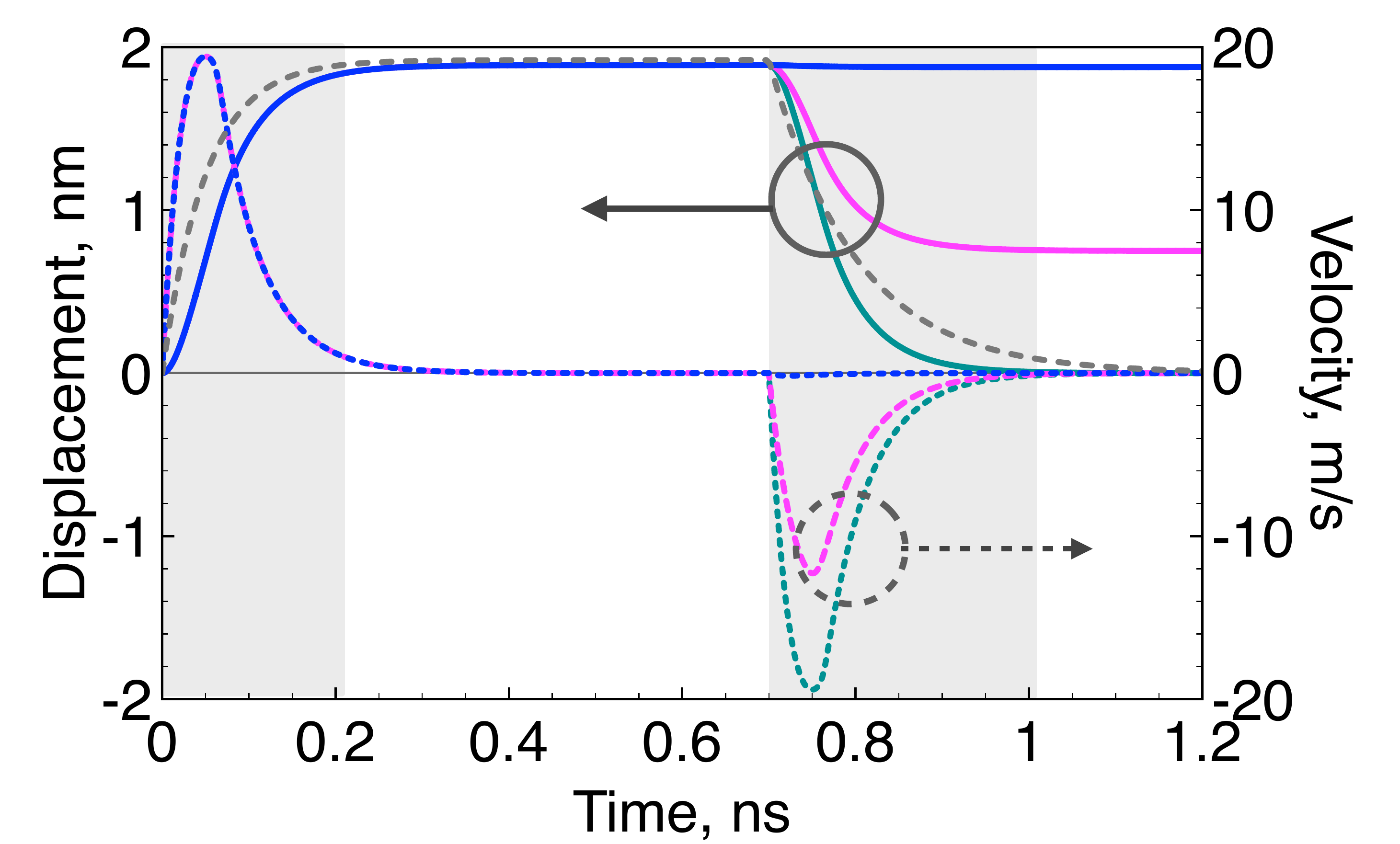}
	\caption[fig_single_pulse]{Time dependence of DW displacement (left axis, solid lines) and velocity (right axis, dashed lines) induced by field pulses with different fall times. $\tau_\mathrm{raise}=$50~ps. Fall time is 50 ps (green line), 100 ps (magenta line), and 200 ps (blue line). Relaxation time $\tau_\mathrm{relax}=$50 ps. The grey dotted line shows the pulse shape with exponential rise/fall $\propto \exp(-t/\tau)$. The rising/falling interval is shown with shaded area.}
	\vspace{-0.4 cm}
	\label{fig_asymmetric_pulse}
\end{figure}

  A more experimentally realistic alternating (cos-like) field $H_\mathrm{ac}\propto\cos(\omega t)$ can only induce  oscillations of the DW  with zero permanent displacement by drift.  The green line in  Fig.\ref{fig_asymmetric_pulse} (left axis) shows the displacement of the DW induced by a symmetric field pulse. {For all  pulses we have taken the time between rise and fall times to be 700 ps and field amplitude $H_\mathrm{ac}$=10 mT}.
  During the rising edge and falling edge periods of the pulses, the DW moves in opposite directions with exactly the same velocity (Fig.\ref{fig_asymmetric_pulse}(right axis)),
  resulting in zero displacement.

Nonzero displacement can be achieved with an asymmetric pulse, as illustrated by the magenta (fall time 100 ps) and blue (fall time 200 ps)  lines in Fig.\ref{fig_asymmetric_pulse}. The corresponding asymmetry of the velocity during the raising and falling intervals (Fig.\ref{fig_asymmetric_pulse} (b)) is due to the frictional force.  Friction sets a threshold for the DW depinning and prevents DW motion for small field rates $\dot{H}_\mathrm{ac}$. As a result,  the velocity of backward motion diminishes with increasing fall time. At some critical value of fall time the backward motion of the DW is blocked (blue lines in Fig.\ref{fig_asymmetric_pulse}) and the displacement of the DW is maximal.
 
The maximal DW displacement during the pulse depends upon the relation between risetime $\tau_\mathrm{raise}$ and relaxation time of the DW, $\tau_\mathrm{relax}$. 
For a given material (fixed relaxation time) the optimal rising time is close to $\tau_{\rm relax}$. For a given experimental technique (fixed raise time) a longer relaxation time is preferable (magenta vs blue lines). Note, that a small relaxation time is typical for the metallic systems like Mn$_2$Au, while a large $\tau_\mathrm{relax}$ is more typical for insulators like NiO. 

\begin{figure}[h]
	\centering
	\includegraphics[width=0.7\columnwidth]{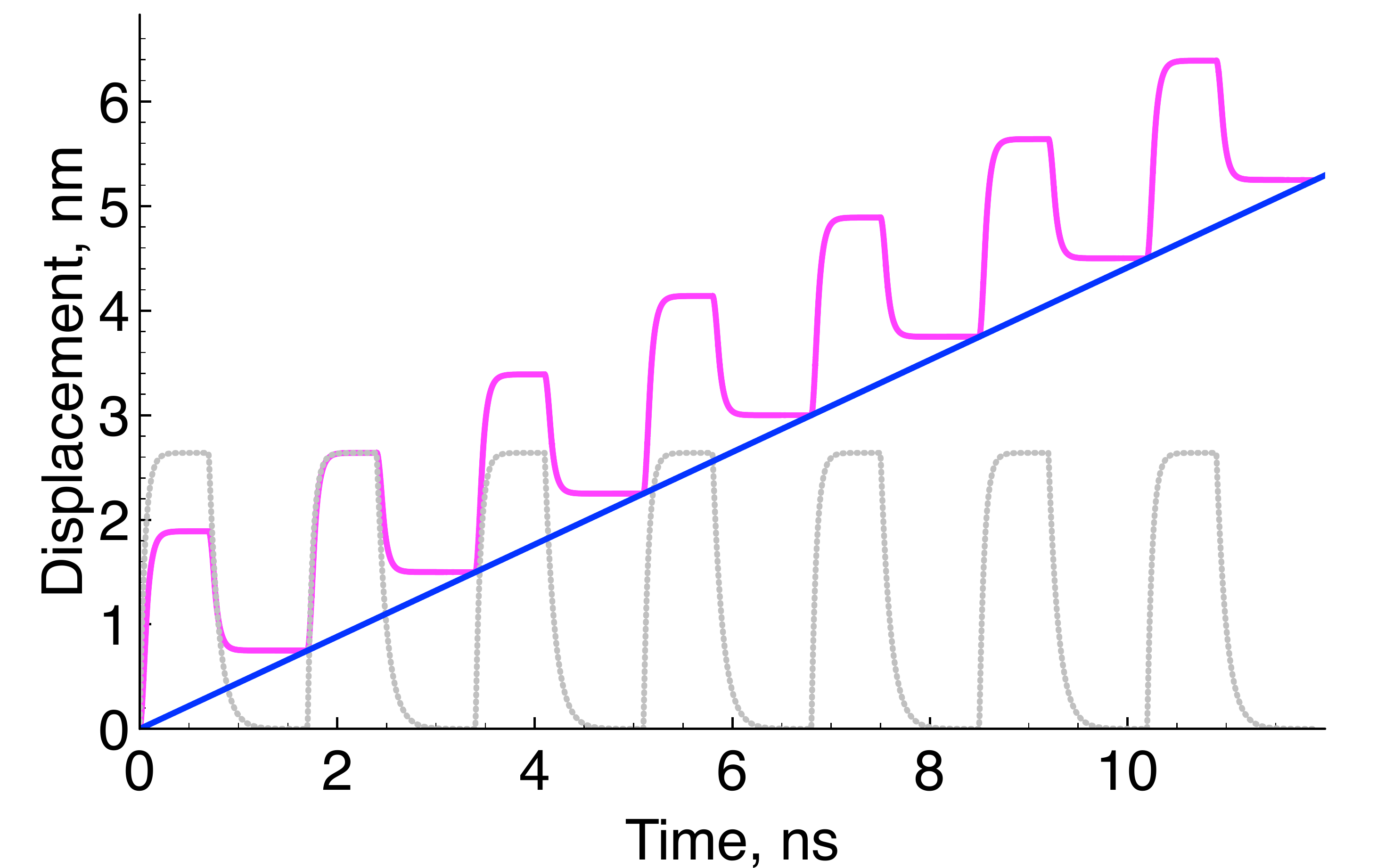}
	\caption[fig_single_pulse]{ Rachet-like displacement (magenta line) of the DW  under the sawtooth-shaped pulses (grey dotted line). Blue line shows the effective displacement as a function of time, average velocity being 0.44 m/s.  $\tau_\mathrm{relax}=\tau_\mathrm{raise}$=50 ps.
		The  fall time is 100 ps.}
	\label{fig_rachet}
\end{figure}

Although the displacement of a DW during one pulse is limited by the internal damping, the friction, and the attainable pulse parameters, a DW can be moved to any distance by a periodic set of pulses, as shown in
Fig.~\ref{fig_rachet} (b). The average velocity of such rachet-like motion  (in this example is 0.44 m/s), can be controlled by a proper choice of the pulse duration and the interval between the pulses. 
To attain maximal velocity, the time between rise and fall times should be  minimized (white range in Fig.~\ref{fig_rachet}).
We also note that this type of ratchet force is different from its counterpart in FM materials, where an oscillating motion is induced instead.\cite{Kruger2014}


In summary, we exploit the use of asymmetric field pulses to displace 
AF DWs. We ascertain that asymmetric sawtooth-shaped pulses of the magnetic field in combination with the natural defect-induced static friction enable unidirectional 
controlled rachet-like motion of an AF DW. This mechanism is broadly applicable to many different types of AF materials and can induced synchronous motion of multiple domain walls as required for applications.

{We acknowledge} support from Humboldt Foundation, the EU (Wall PEOPLE-2013-ITN 608031;  MultiRev ERC-2014-PoC 665672) as well as the Center of Innovative and Emerging Materials at Johannes Gutenberg University Mainz, the Graduate School
of Excellence Materials Science in Mainz (CSC 266),
the DFG (in particular SFB TRR 173 Spin+X), the Ministry of Education of the Czech Republic Grant No. LM2011026, and from the Grant Agency of the Czech Republic Grant no. 14-37427


\begin{thebibliography}{18}%
	\makeatletter
	\providecommand \@ifxundefined [1]{%
		\@ifx{#1\undefined}
	}%
	\providecommand \@ifnum [1]{%
		\ifnum #1\expandafter \@firstoftwo
		\else \expandafter \@secondoftwo
		\fi
	}%
	\providecommand \@ifx [1]{%
		\ifx #1\expandafter \@firstoftwo
		\else \expandafter \@secondoftwo
		\fi
	}%
	\providecommand \natexlab [1]{#1}%
	\providecommand \enquote  [1]{``#1''}%
	\providecommand \bibnamefont  [1]{#1}%
	\providecommand \bibfnamefont [1]{#1}%
	\providecommand \citenamefont [1]{#1}%
	\providecommand \href@noop [0]{\@secondoftwo}%
	\providecommand \href [0]{\begingroup \@sanitize@url \@href}%
	\providecommand \@href[1]{\@@startlink{#1}\@@href}%
	\providecommand \@@href[1]{\endgroup#1\@@endlink}%
	\providecommand \@sanitize@url [0]{\catcode `\\12\catcode `\$12\catcode
		`\&12\catcode `\#12\catcode `\^12\catcode `\_12\catcode `\%12\relax}%
	\providecommand \@@startlink[1]{}%
	\providecommand \@@endlink[0]{}%
	\providecommand \url  [0]{\begingroup\@sanitize@url \@url }%
	\providecommand \@url [1]{\endgroup\@href {#1}{\urlprefix }}%
	\providecommand \urlprefix  [0]{URL }%
	\providecommand \Eprint [0]{\href }%
	\providecommand \doibase [0]{http://dx.doi.org/}%
	\providecommand \selectlanguage [0]{\@gobble}%
	\providecommand \bibinfo  [0]{\@secondoftwo}%
	\providecommand \bibfield  [0]{\@secondoftwo}%
	\providecommand \translation [1]{[#1]}%
	\providecommand \BibitemOpen [0]{}%
	\providecommand \bibitemStop [0]{}%
	\providecommand \bibitemNoStop [0]{.\EOS\space}%
	\providecommand \EOS [0]{\spacefactor3000\relax}%
	\providecommand \BibitemShut  [1]{\csname bibitem#1\endcsname}%
	\let\auto@bib@innerbib\@empty
	\bibitem [{\citenamefont {MacDonald}\ and\ \citenamefont
		{Tsoi}(2011)}]{MacDonald2011}%
	\BibitemOpen
	\bibfield  {author} {\bibinfo {author} {\bibfnamefont {A.~H.}\ \bibnamefont
			{MacDonald}}\ and\ \bibinfo {author} {\bibfnamefont {M.}~\bibnamefont
			{Tsoi}},\ }\bibfield  {title} {\enquote {\bibinfo {title} {{Antiferromagnetic
					metal spintronics.}}}\ }\href {\doibase 10.1098/rsta.2011.0014} {\bibfield
		{journal} {\bibinfo  {journal} {Phil. Trans. R. Soc. A }\ }\textbf {\bibinfo
			{volume} {369}},\ \bibinfo {pages} {3098} (\bibinfo {year}
		{2011})}\BibitemShut {NoStop}%
	\bibitem [{\citenamefont {Gomonay}\ and\ \citenamefont
		{Loktev}(2014)}]{Gomonay2014}%
	\BibitemOpen
	\bibfield  {author} {\bibinfo {author} {\bibfnamefont {E.~V.}\ \bibnamefont
			{Gomonay}}\ and\ \bibinfo {author} {\bibfnamefont {V.~M.}\ \bibnamefont
			{Loktev}},\ }\bibfield  {title} {\enquote {\bibinfo {title} {{Spintronics of
					antiferromagnetic systems}},}\ }\href {\doibase
		10.1063/1.4862467} {\bibfield  {journal} {\bibinfo  {journal} {Low
				Temp. Phys.}\ }\textbf {\bibinfo {volume} {40}},\ \bibinfo {pages}
		{17} (\bibinfo {year} {2014})}\BibitemShut {NoStop}%
	\bibitem [{\citenamefont {Jungwirth}\ \emph {et~al.}(2016)\citenamefont
		{Jungwirth}, \citenamefont {Marti}, \citenamefont {Wadley},\ and\
		\citenamefont {Wunderlich}}]{Jungwirth2016}%
	\BibitemOpen
	\bibfield  {author} {\bibinfo {author} {\bibfnamefont {T.}~\bibnamefont
			{Jungwirth}}, \bibinfo {author} {\bibfnamefont {X.}~\bibnamefont {Marti}},
		\bibinfo {author} {\bibfnamefont {P.}~\bibnamefont {Wadley}}, \ and\ \bibinfo
		{author} {\bibfnamefont {J.}~\bibnamefont {Wunderlich}},\ }\bibfield  {title}
	{\enquote {\bibinfo {title} {{Antiferromagnetic spintronics}},}\ }\href
	{\doibase 10.1038/nnano.2016.18} {\bibfield  {journal} {\bibinfo  {journal}
			{Nat. Nanotechnol.}\ }\textbf {\bibinfo {volume} {11}},\ \bibinfo {pages}
		{231} (\bibinfo {year} {2016})}\BibitemShut {NoStop}%
	\bibitem [{\citenamefont {Yang}, \citenamefont {Ryu},\ and\ \citenamefont
		{Parkin}(2015)}]{Yang2015a}%
	\BibitemOpen
	\bibfield  {author} {\bibinfo {author} {\bibfnamefont {S.-H.}\ \bibnamefont
			{Yang}}, \bibinfo {author} {\bibfnamefont {K.-S.}\ \bibnamefont {Ryu}}, \
		and\ \bibinfo {author} {\bibfnamefont {S.}~\bibnamefont {Parkin}},\
	}\bibfield  {title} {\enquote {\bibinfo {title} {{Domain-wall velocities of
				up to 750 m s−1 driven by exchange-coupling torque in synthetic
				antiferromagnets}},}\ }\href {\doibase 10.1038/nnano.2014.324} {\bibfield
	{journal} {\bibinfo  {journal} {Nat. Nanotechnol.}\ }\textbf {\bibinfo
		{volume} {10}},\ \bibinfo {pages} {221} (\bibinfo {year}
	{2015})}\BibitemShut {NoStop}%
\bibitem [{\citenamefont {Barthem}\ \emph {et~al.}(2016)\citenamefont
	{Barthem}, \citenamefont {Colin}, \citenamefont {Haettel}, \citenamefont
	{Dufeu},\ and\ \citenamefont {Givord}}]{Barthem2016}%
\BibitemOpen
\bibfield  {author} {\bibinfo {author} {\bibfnamefont {V.~M. T.~S.}\
		\bibnamefont {Barthem}}, \bibinfo {author} {\bibfnamefont {C.~V.}\
		\bibnamefont {Colin}}, \bibinfo {author} {\bibfnamefont {R.}~\bibnamefont
		{Haettel}}, \bibinfo {author} {\bibfnamefont {D.}~\bibnamefont {Dufeu}}, \
	and\ \bibinfo {author} {\bibfnamefont {D.}~\bibnamefont {Givord}},\
}\bibfield  {title} {\enquote {\bibinfo {title} {{Easy moment direction and
			antiferromagnetic domain wall motion in Mn2Au}},}\ }\href {\doibase
10.1016/j.jmmm.2015.07.101} {\bibfield  {journal} {\bibinfo  {journal}
	{J. Magn. Magn. Mater.}\ }\textbf {\bibinfo {volume}
	{406}},\ \bibinfo {pages} {289} (\bibinfo {year} {2016})}\BibitemShut
{NoStop}%
\bibitem [{\citenamefont {Gomonay}, \citenamefont {Jungwirth},\ and\
	\citenamefont {Sinova}(2016)}]{Gomonay2016}%
\BibitemOpen
\bibfield  {author} {\bibinfo {author} {\bibfnamefont {O.}~\bibnamefont
		{Gomonay}}, \bibinfo {author} {\bibfnamefont {T.}~\bibnamefont {Jungwirth}},
	\ and\ \bibinfo {author} {\bibfnamefont {J.}~\bibnamefont {Sinova}},\
}\bibfield  {title} {\enquote {\bibinfo {title} {{High
			antiferromagnetic domain wall velocity in a staggered spin-orbit field}},}\
}\href {\doibase 10.1103/PhysRevLett.117.017202} {\bibfield  {journal}
{\bibinfo  {journal} {Phys. Rev. Lett.}\ }\textbf {\bibinfo {volume} {117}},\
\bibinfo {pages} {017202} (\bibinfo {year} {2016})}\BibitemShut {NoStop}%
\bibitem [{\citenamefont {Shiino}\ \emph {et~al.}(2016)\citenamefont {Shiino},
	\citenamefont {Oh}, \citenamefont {Haney}, \citenamefont {Lee}, \citenamefont
	{Go}, \citenamefont {Park},\ and\ \citenamefont {Lee}}]{Shiino2016}%
\BibitemOpen
\bibfield  {author} {\bibinfo {author} {\bibfnamefont {T.}~\bibnamefont
		{Shiino}}, \bibinfo {author} {\bibfnamefont {S.-H.}\ \bibnamefont {Oh}},
	\bibinfo {author} {\bibfnamefont {P.~M.}\ \bibnamefont {Haney}}, \bibinfo
	{author} {\bibfnamefont {S.-W.}\ \bibnamefont {Lee}}, \bibinfo {author}
	{\bibfnamefont {G.}~\bibnamefont {Go}}, \bibinfo {author} {\bibfnamefont
		{B.-G.}\ \bibnamefont {Park}}, \ and\ \bibinfo {author} {\bibfnamefont
		{K.-J.}\ \bibnamefont {Lee}},\ }\bibfield  {title} {\enquote {\bibinfo
		{title} {{Antiferromagnetic domain wall motion driven by spin-orbit
				torques}},}\ }\href {\doibase 10.1103/PhysRevLett.117.087203} {\bibfield  {journal}
	{\bibinfo  {journal} {Phys. Rev. Lett.}\ }\textbf {\bibinfo {volume} {117}},\
	\bibinfo {pages} {087203} (\bibinfo {year} {2016})}
\BibitemShut {NoStop}%
\bibitem [{\citenamefont {Kim}, \citenamefont {Tchernyshyov},\ and\
	\citenamefont {Tserkovnyak}(2015)}]{Kim2015d}%
\BibitemOpen
\bibfield  {author} {\bibinfo {author} {\bibfnamefont {S.~K.}\ \bibnamefont
		{Kim}}, \bibinfo {author} {\bibfnamefont {O.}~\bibnamefont {Tchernyshyov}}, \
	and\ \bibinfo {author} {\bibfnamefont {Y.}~\bibnamefont {Tserkovnyak}},\
}\bibfield  {title} {\enquote {\bibinfo {title} {{Thermophoresis of an
			antiferromagnetic soliton}},}\ }\href {\doibase 10.1103/PhysRevB.92.020402}
{\bibfield  {journal} {\bibinfo  {journal} {Phys. Rev. B}\ }\textbf {\bibinfo
		{volume} {92}},\ \bibinfo {pages} {020402} (\bibinfo {year} {2015})}
\BibitemShut {NoStop}%
\bibitem [{\citenamefont {Selzer}\ \emph {et~al.}(2016)\citenamefont {Selzer},
	\citenamefont {Atxitia}, \citenamefont {Ritzmann}, \citenamefont {Hinzke},\
	and\ \citenamefont {Nowak}}]{Selzer2016}%
\BibitemOpen
\bibfield  {author} {\bibinfo {author} {\bibfnamefont {S.}~\bibnamefont
		{Selzer}}, \bibinfo {author} {\bibfnamefont {U.}~\bibnamefont {Atxitia}},
	\bibinfo {author} {\bibfnamefont {U.}~\bibnamefont {Ritzmann}}, \bibinfo
	{author} {\bibfnamefont {D.}~\bibnamefont {Hinzke}}, \ and\ \bibinfo {author}
	{\bibfnamefont {U.}~\bibnamefont {Nowak}},\ }\href@noop {} {\enquote
	{\bibinfo {title} {{Inertia-free Thermally Driven Domain Wall Motion in
				Antiferromagnets}},}\ } 
{\bibfield  {journal} {\bibinfo  {journal} {Phys. Rev. Lett.}\ } (\bibinfo {year} {2016})}\BibitemShut {NoStop}%
\bibitem [{\citenamefont {Haldane}(1983)}]{Haldane1983a}%
\BibitemOpen
\bibfield  {author} {\bibinfo {author} {\bibfnamefont {F.~D.~M.}\
		\bibnamefont {Haldane}},\ }\bibfield  {title} {\enquote {\bibinfo {title}
		{{Nonlinear field theory of large-spin Heisenberg antiferromagnets:
				Semiclassically quantized solitons of the one-dimensional easy-axis Neel
				state}},}\ }\href {\doibase 10.1103/PhysRevLett.50.1153} {\bibfield
	{journal} {\bibinfo  {journal} {Phys. Rev. Lett.}\ }\textbf {\bibinfo
		{volume} {50}},\ \bibinfo {pages} {1153} (\bibinfo {year}
	{1983})}\BibitemShut {NoStop}%
\bibitem [{\citenamefont {Kosevich}, \citenamefont {Ivanov},\ and\
	\citenamefont {Kovalev}(1990)}]{Kosevich1990}%
\BibitemOpen
\bibfield  {author} {\bibinfo {author} {\bibfnamefont {A.}~\bibnamefont
		{Kosevich}}, \bibinfo {author} {\bibfnamefont {B.}~\bibnamefont {Ivanov}}, \
	and\ \bibinfo {author} {\bibfnamefont {A.}~\bibnamefont {Kovalev}},\
}\bibfield  {title} {\enquote {\bibinfo {title} {{Magnetic Solitons}},}\
}\href {\doibase 10.1016/0370-1573(90)90130-T} {\bibfield  {journal}
{\bibinfo  {journal} {Physics Reports}\ }\textbf {\bibinfo {volume} {194}},\
\bibinfo {pages} {117} (\bibinfo {year} {1990})}\BibitemShut {NoStop}%
\bibitem [{\citenamefont {Ivanov}\ and\ \citenamefont
	{Kolezhuk}(1995)}]{Ivanov1995}%
\BibitemOpen
\bibfield  {author} {\bibinfo {author} {\bibfnamefont {B.~A.}\ \bibnamefont
		{Ivanov}}\ and\ \bibinfo {author} {\bibfnamefont {A.~K.}\ \bibnamefont
		{Kolezhuk}},\ }\bibfield  {title} {\enquote {\bibinfo {title} {{Solitons in
				low-dimensional antiferromagnets}},}\
}\href@noop {} {\bibfield  {journal} {\bibinfo  {journal} {Low Temp. Phys.
		}\ }\textbf {\bibinfo {volume} {21}},\ \bibinfo {pages}
{275} (\bibinfo {year} {1995})}\BibitemShut {NoStop}%
\bibitem [{\citenamefont {Gomonay}\ and\ \citenamefont
	{Loktev}(2010)}]{Gomonay2010}%
\BibitemOpen
\bibfield  {author} {\bibinfo {author} {\bibfnamefont {H.~V.}\ \bibnamefont
		{Gomonay}}\ and\ \bibinfo {author} {\bibfnamefont {V.~M.}\ \bibnamefont
		{Loktev}},\ }\bibfield  {title} {\enquote {\bibinfo {title} {{Spin transfer
				and current-induced switching in antiferromagnets}},}\ }\href {\doibase
	10.1103/PhysRevB.81.144427} {\bibfield  {journal} {\bibinfo  {journal} {Phys.
			Rev. B}\ }\textbf {\bibinfo {volume} {81}},\ \bibinfo {pages} {144427}
	(\bibinfo {year} {2010})}\BibitemShut {NoStop}%
\bibitem [{\citenamefont {Wu}\ \emph {et~al.}(2012)\citenamefont {Wu},
	\citenamefont {Liao}, \citenamefont {Sofin}, \citenamefont {Feng},
	\citenamefont {Ma}, \citenamefont {Shick}, \citenamefont {Mryasov},\ and\
	\citenamefont {Shvets}}]{Wu2012}%
\BibitemOpen
\bibfield  {author} {\bibinfo {author} {\bibfnamefont {H.~C.}\ \bibnamefont
		{Wu}}, \bibinfo {author} {\bibfnamefont {Z.~M.}\ \bibnamefont {Liao}},
	\bibinfo {author} {\bibfnamefont {R.~G.~S.}\ \bibnamefont {Sofin}}, \bibinfo
	{author} {\bibfnamefont {G.}~\bibnamefont {Feng}}, \bibinfo {author}
	{\bibfnamefont {X.~M.}\ \bibnamefont {Ma}}, \bibinfo {author} {\bibfnamefont
		{A.~B.}\ \bibnamefont {Shick}}, \bibinfo {author} {\bibfnamefont {O.~N.}\
		\bibnamefont {Mryasov}}, \ and\ \bibinfo {author} {\bibfnamefont {I.~V.}\
		\bibnamefont {Shvets}},\ }\bibfield  {title} {\enquote {\bibinfo {title}
		{{Mn2Au: Body-centered-tetragonal bimetallic antiferromagnets grown by
				molecular beam epitaxy}},}\ }\href {\doibase 10.1002/adma.201202273}
{\bibfield  {journal} {\bibinfo  {journal} {Advanced Materials}\ }\textbf
	{\bibinfo {volume} {24}},\ \bibinfo {pages} {6374} (\bibinfo {year}
	{2012})}\BibitemShut {NoStop}%
\bibitem [{\citenamefont {Shick}\ \emph {et~al.}(2010)\citenamefont {Shick},
	\citenamefont {Khmelevskyi}, \citenamefont {Mryasov}, \citenamefont
	{Wunderlich},\ and\ \citenamefont {Jungwirth}}]{Shick2010}%
\BibitemOpen
\bibfield  {author} {\bibinfo {author} {\bibfnamefont {A.~B.}\ \bibnamefont
		{Shick}}, \bibinfo {author} {\bibfnamefont {S.}~\bibnamefont {Khmelevskyi}},
	\bibinfo {author} {\bibfnamefont {O.~N.}\ \bibnamefont {Mryasov}}, \bibinfo
	{author} {\bibfnamefont {J.}~\bibnamefont {Wunderlich}}, \ and\ \bibinfo
	{author} {\bibfnamefont {T.}~\bibnamefont {Jungwirth}},\ }\bibfield  {title}
{\enquote {\bibinfo {title} {{Spin-orbit coupling induced anisotropy effects
				in bimetallic antiferromagnets: A route towards antiferromagnetic
				spintronics}},}\ }\href {\doibase 10.1103/PhysRevB.81.212409} {\bibfield
	{journal} {\bibinfo  {journal} {Phys. Rev. B}\ }\textbf {\bibinfo {volume}
		{81}},\ \bibinfo {pages} {212409} (\bibinfo {year} {2010})}\BibitemShut
{NoStop}%
\bibitem [{\citenamefont {Hutchings}\ and\ \citenamefont
	{Samuelsen}(1972)}]{Hutchings1972}%
\BibitemOpen
\bibfield  {author} {\bibinfo {author} {\bibfnamefont {M.~T.}\ \bibnamefont
		{Hutchings}}\ and\ \bibinfo {author} {\bibfnamefont {E.~J.}\ \bibnamefont
		{Samuelsen}},\ }\bibfield  {title} {\enquote {\bibinfo {title} {{Measurement
				of Spin-Wave Dispersion in NiO by Inelastic Neutron Scattering and Its
				Relation to Magnetic Properties}},}\ }\href {\doibase
	10.1103/PhysRevB.6.3447} {\bibfield  {journal} {\bibinfo  {journal} {Phys.
			Rev. B}\ }\textbf {\bibinfo {volume} {6}},\ \bibinfo {pages} {3447}
	(\bibinfo {year} {1972})}\BibitemShut {NoStop}%
\bibitem [{\citenamefont {Kampfrath}\ \emph {et~al.}(2010)\citenamefont
	{Kampfrath}, \citenamefont {Sell}, \citenamefont {Klatt}, \citenamefont
	{Pashkin}, \citenamefont {M{\"{a}}hrlein}, \citenamefont {Dekorsy},
	\citenamefont {Wolf}, \citenamefont {Fiebig}, \citenamefont {Leitenstorfer},\
	and\ \citenamefont {Huber}}]{Kampfrath2010}%
\BibitemOpen
\bibfield  {author} {\bibinfo {author} {\bibfnamefont {T.}~\bibnamefont
		{Kampfrath}}, \bibinfo {author} {\bibfnamefont {A.}~\bibnamefont {Sell}},
	\bibinfo {author} {\bibfnamefont {G.}~\bibnamefont {Klatt}}, \bibinfo
	{author} {\bibfnamefont {A.}~\bibnamefont {Pashkin}}, \bibinfo {author}
	{\bibfnamefont {S.}~\bibnamefont {M{\"{a}}hrlein}}, \bibinfo {author}
	{\bibfnamefont {T.}~\bibnamefont {Dekorsy}}, \bibinfo {author} {\bibfnamefont
		{M.}~\bibnamefont {Wolf}}, \bibinfo {author} {\bibfnamefont {M.}~\bibnamefont
		{Fiebig}}, \bibinfo {author} {\bibfnamefont {A.}~\bibnamefont
		{Leitenstorfer}}, \ and\ \bibinfo {author} {\bibfnamefont {R.}~\bibnamefont
		{Huber}},\ }\bibfield  {title} {\enquote {\bibinfo {title} {{Coherent
				terahertz control of antiferromagnetic spin waves}},}\ }\href {\doibase
	10.1038/nphoton.2010.259} {\bibfield  {journal} {\bibinfo  {journal} {Nat.
			Photonics}\ }\textbf {\bibinfo {volume} {5}},\ \bibinfo {pages} {31--34}
	(\bibinfo {year} {2010})}\BibitemShut {NoStop}%
\bibitem [{\citenamefont {Kim}\ \emph {et~al.}(2014)\citenamefont {Kim},
	\citenamefont {Mawass}, \citenamefont {Bisig}, \citenamefont {Kr{\"{u}}ger},
	\citenamefont {Reeve}, \citenamefont {Schulz}, \citenamefont {B{\"{u}}ttner},
	\citenamefont {Yoon}, \citenamefont {You}, \citenamefont {Weigand},
	\citenamefont {Stoll}, \citenamefont {Sch{\"{u}}tz}, \citenamefont {Swagten},
	\citenamefont {Koopmans}, \citenamefont {Eisebitt},\ and\ \citenamefont
	{Kl{\"{a}}ui}}]{Kruger2014}%
\BibitemOpen
\bibfield  {author} {\bibinfo {author} {\bibfnamefont {J.-s.}\ \bibnamefont
		{Kim}}, \bibinfo {author} {\bibfnamefont {M.-A.}\ \bibnamefont {Mawass}},
	\bibinfo {author} {\bibfnamefont {A.}~\bibnamefont {Bisig}}, \bibinfo
	{author} {\bibfnamefont {B.}~\bibnamefont {Kr{\"{u}}ger}}, \bibinfo {author}
	{\bibfnamefont {R.~M.}\ \bibnamefont {Reeve}}, \bibinfo {author}
	{\bibfnamefont {T.}~\bibnamefont {Schulz}}, \bibinfo {author} {\bibfnamefont
		{F.}~\bibnamefont {B{\"{u}}ttner}}, \bibinfo {author} {\bibfnamefont
		{J.}~\bibnamefont {Yoon}}, \bibinfo {author} {\bibfnamefont {C.-y.}\
		\bibnamefont {You}}, \bibinfo {author} {\bibfnamefont {M.}~\bibnamefont
		{Weigand}}, \bibinfo {author} {\bibfnamefont {H.}~\bibnamefont {Stoll}},
	\bibinfo {author} {\bibfnamefont {G.}~\bibnamefont {Sch{\"{u}}tz}}, \bibinfo
	{author} {\bibfnamefont {H.~J.~M.}\ \bibnamefont {Swagten}}, \bibinfo
	{author} {\bibfnamefont {B.}~\bibnamefont {Koopmans}}, \bibinfo {author}
	{\bibfnamefont {S.}~\bibnamefont {Eisebitt}}, \ and\ \bibinfo {author}
	{\bibfnamefont {M.}~\bibnamefont {Kl{\"{a}}ui}},\ }\bibfield  {title}
{\enquote {\bibinfo {title} {{Synchronous precessional motion of multiple
				domain walls in a ferromagnetic nanowire by perpendicular field pulses}},}\
}\href {\doibase 10.1038/ncomms4429} {\bibfield  {journal} {\bibinfo
	{journal} {Nat. Comm.}\ }\textbf {\bibinfo {volume} {5}},\
\bibinfo {pages} {3492} (\bibinfo {year} {2014})}\BibitemShut {NoStop}%
\end{thebibliography}
%

\end{document}